\documentclass{article}
\usepackage[T1]{fontenc}
\usepackage{titlesec, blindtext, color}
\definecolor{gray75}{gray}{0.75}
\newcommand{\hsp}{\hspace{20pt}}
\titleformat{\chapter}[hang]{\Huge\bfseries}{\thechapter\hsp\textcolor{gray75}{|}\hsp}{0pt}{\Huge\bfseries}

\usepackage[margin=1in]{geometry}
\usepackage{bm}
\usepackage[english]{babel}
\usepackage[latin1]{inputenc}
\usepackage{times}
\usepackage{graphics,graphicx,pgf,natbib,amsmath,amsthm,amssymb,url}
\usepackage{tgbonum}
\usepackage{rotating}
\usepackage{tfrupee}
\usepackage{natbib}
\usepackage{tfrupee} 
\usepackage{algorithm2e}

\usepackage{hyperref}
\usepackage{tcolorbox}
\usepackage{tikz}
\usetikzlibrary{"arrows", "automata", "backgrounds", "calendar", "chains", "matrix", "mindmap", "patterns", "petri", "shadows", "shapes.geometric", "shapes.misc", "spy", "trees"}

\newcommand{\bfb} {\mbox{\boldmath $\beta$}}

\newcommand{\bfo} {\mbox{\boldmath $\omega$}}

\newcommand{\bfe} {\mbox{\boldmath $\epsilon$}}
\newcommand{\bfep} {\mbox{\boldmath $\varepsilon$}}

\newcommand{\bfvr} {\mbox{\boldmath $\varrho$}}

\newcommand{\bfm} {\mbox{\boldmath $\mu$}}
\newcommand{\bfS} {\mbox{\boldmath $\Sigma$}}

\newcommand{\bfP} {\mbox{\boldmath $\Psi$}}

\newcommand{\I}{{\mbox{\boldmath $I$}}}

\newcommand{\x}{{\mbox{\boldmath $x$}}}
\newcommand{\rb}{{\mbox{\boldmath $r$}}}
\newcommand{\X}{{\mbox{\boldmath $X$}}}

\theoremstyle{plain}
\numberwithin{equation}{section}
\newtheorem{Defn}{Definition}[section]
\newtheorem{theorem}{Theorem}[section]
\newtheorem{result}{Result}[section]
\newtheorem{remark}{Remark}[section]
\newtheorem{example}{Example}[section]

\title{\textbf{Risk Analysis of Passive Portfolios }}
\author{
    Sourish Das\\
    Chennai Mathematical Institute
}
\date{July 2024}

\begin{document}
\maketitle

\begin{abstract}
In this work, we present an alternative passive investment strategy. The passive investment philosophy comes from the Efficient Market Hypothesis (EMH), and its adoption is widespread. If EMH is true, one cannot outperform market by actively managing their portfolio for a long time. Also, it requires little to no intervention. People can buy an exchange-traded fund (ETF) with a long-term perspective. As the economy grows over time, one expects the ETF to grow. For example, in India, one can invest in NETF (see,\cite{NETF}), which suppose to mimic the Nifty50 return. However, the weights of the Nifty 50 index are based on market capitalisation. These weights are not necessarily optimal for the investor. In this work, we present that volatility risk and extreme risk measures of the Nifty50 portfolio are uniformly larger than Markowitz's optimal portfolio. However, common people can't create an optimised portfolio. So we proposed an alternative passive investment strategy of an equal-weight portfolio. We show that if one pushes the maximum weight of the portfolio towards equal weight, the idiosyncratic risk of the portfolio would be minimal. The empirical evidence indicates that the risk profile of an equal-weight portfolio is similar to that of Markowitz's optimal portfolio. Hence instead of buying Nifty50 ETFs, one should equally invest in the stocks of Nifty50 to achieve a uniformly better risk profile than the Nifty 50 ETF portfolio. We also present an analysis of how portfolios perform to idiosyncratic events like the Russian invasion of Ukraine. We found that the equal weight portfolio has a uniformly lower risk than the Nifty 50 portfolio before and during the Russia-Ukraine war. All codes are available on GitHub (\url{https://github.com/sourish-cmi/quant/tree/main/Chap_Risk_Anal_of_Passive_Portfolio}).
\end{abstract}

\section{Introduction}

In recent years, we observed several undesirable events, such as the Covid19 pandemic, the Russian invasion of Ukraine, severe global supply chain system disruption, and tremendous pressure on energy prices, followed by global inflation. These macro events bring a high degree of uncertainty to the worldwide market. Nowadays, even pension fund portfolios invest in the volatile equity market. The high degree of volatility can impact the portfolio negatively. Therefore it is essential to analyse the risk exposure that a portfolio face.

Passive investment is an excellent option if the investment horizon is long and one is not an expert in quantitative finance. The widespread adoption of the passive investment philosophy comes from the Efficient Market Hypothesis (EMH). One cannot outperform the market by actively managing their portfolio for a long time if it is efficient. One can invest in exchange-traded funds (ETF) with a long-term perspective and expects the ETF to grow as the economy grows over time. In India, we can invest in NETF (see, \cite{NETF}), which suppose to mimic the Nifty50 return. However, the weights of the Nifty 50 index are not necessarily optimal for the investors, as they are based on market capitalisation. 

In this work, we showed that the Indian market is not efficient. It raises the question of how it affects the risk profile of the passive investment portfolio, particularly how it reacts to idiosyncratic events like the Russian invasion of Ukraine. We study the effect of Russian invasion on the risk profile of the portfolios. 

Rest of the chapter is organised as follows. In Section (\ref{sec_basic_finance}) we present basics of financial return and volatility risk. In Section (\ref{sec_EMH}) we present the efficient market hypothesis and empirical evidence that shows Indian market is not efficient. In Section (\ref{sec_portfolio_risk_analysis}) we present different aspects of portfolio risk analysis. In Section (\ref{sec_empirical_evidence}), we present the empirical evidence that equal weight portfolio would have uniformly better  risk profile than Nifty50 ETFs. Section (\ref{sec_conclusion}) concludes the chapter.

\section{On Basics of Financial Return and Volatility Risk }\label{sec_basic_finance}
\subsection{Time Value of Money}
If given a choice between receiving \rupee 100 today or after a year, we should choose to receive \rupee 100 today.  Because if a bank (say State Bank of India) agrees to pay us \rupee 7 for keeping the \rupee  100 with them, then at the end of the period, our investment will be \rupee 107. This \rupee 7 is the time value of \rupee  100 that we are going to keep with the bank. Economists term this as \emph{time preference}, also known as the `\textbf{Time Value of Money}.'  The mathematical operation of evaluating the `present value' (PV)  of an amount into the `future value' (FV) is called a \textbf{capitalization}. For example, how much will our \rupee 100 today be worth in 10 years? The reverse operation of evaluating the present value of a future amount of money is called the \textbf{discounting}. For example, how much will \rupee 100 received in 10 years, be worth today?

\subsection{Financial Return}
The goal of our any investment is to grow over time. The growth depends on both change in price and a number of the assets being invested. Certainly, our would be interested in revenues that are high compared to the initial investment. Returns articulate the change in price as a fraction of the initial price.

\noindent Suppose $P_t$ is the price of an asset at time $t$. The \textbf{net return} over the holding period of time $t-1$ to $t$ is
$$
R_t=\frac{P_t-P_{t-1}}{P_{t-1}}=\frac{P_t}{P_{t-1}}-1.
$$
The numerator, $P_t-P_{t-1}$ is the net profit (or net loss) during the holding period, where denominator $P_{t-1}$ is the initial investment at the start of the holding period. We can see the net returns as the rate of profit (or loss) or relative revenue on initial investment. The revenue from holding an asset is 
$$
P_{t}-P_{t-1}=R_t \times P_{t-1},
$$
\begin{center}
revenue = net return $\times$ initial investment.
\end{center}

\begin{tcolorbox}
\begin{example}\label{RFR_ex_1}
An initial investment of \rupee 1000 and a net return of 4\% over one year earn revenue of \rupee 40, which means the value of the investment after one year is \rupee 1040.
\end{example}
\end{tcolorbox}
The single period \textbf{gross return} is defined as
$$
\frac{P_t}{P_{t-1}}=1+R_t
$$

\begin{tcolorbox}
\begin{example}\label{RFR_ex_2}
If $P_{t-1}=$\rupee 1000 and $P_{t}$=\rupee 1040 then the gross return is $1+R_t=1.04$ or $104\%$ and the net return is $R_t=0.04$ or $4\%$.
\end{example}
\end{tcolorbox}

The $k$-period \textbf{gross return} is the product of the $k$ single period, gross returns from time $t-k$ to time $t$:
\begin{eqnarray*}
1+R_t(k)=\frac{P_t}{P_{t-k}}&=&\Big(\frac{P_t}{P_{t-1 }}\Big)\Big(\frac{P_{t-1}}{P_{t-2}}\Big)\hdots\Big(\frac{P_{t-k+1}}{P_{t-k}}\Big)\\
&=& (1+R_t)(1+R_{t-1})\hdots(1+R_{t-k+1}).
\end{eqnarray*}
\begin{tcolorbox}
\begin{example}\label{RFR_ex_3}
Suppose in three consecutive periods (from $t$ to $t+3$) the values of an asset are $P_t$=\rupee  1000, $P_{t+1}$= \rupee 1040, $P_{t+2}$=\rupee 1035, and $P_{t+3}$=\rupee 1050.
$$
1+R_{t+3}(1)=\frac{P_{t+3}}{P_{t+2}}=\frac{1050}{1035}=1.014
$$
$$
1+R_{t+3}(2)=\frac{P_{t+3}}{P_{t+1}}=\frac{1050}{1040}=1.01
$$
$$
1+R_{t+3}(3)=\frac{P_{t+3}}{P_{t}}=\frac{1050}{1000}=1.05
$$
\end{example}
\end{tcolorbox}
We see returns are independent of scale. It does not depend on the unit like rupees, dollar or pounds. However, it depends on the unit of time (like the hour, day, and year).

The \textbf{log returns} are defined as
$$
r_t=\log(1+R_t)=\log(1+R_t)=\log\Big(\frac{P_t}{P_{t-1}}\Big)=p_t-p_{t-1},
$$
where $p_t=\log(P_t)$  is also known as log-price or log-value of the asset. The log-returns are also known as continuously compounded returns.  One advantage of the log returns is that a $k$ period log return is sum of single period log returns. That is
\begin{eqnarray*}
r_t(k)&=&\log\{1+R_t(k)\}\\
&=& \log\{(1+R_t)(1+R_{t-1})...(1+R_{t-k+1})\}\\
&=& \log(1+R_t)+\log(1+R_{t-1})+\hdots+\log(1+R_{t-k+1})\\
&=& r_t+r_{t-1}+\hdots+r_{t-k+1}.
\end{eqnarray*}
We can show, if $x$ is small, then $\log(1+x)\approx x$, it means the log returns are approximately equal to net returns.  Empirical studies show that the condition $\log(1+x)\approx x$ is true when $|x|<0.1$ , i.e., returns that are less than 10\%.
\noindent \textbf{Compounding}\index{Compounding}
If we have an initial investment of $P_t$  that earns annual rate $r$, compounded $m$ times a year for $n$ years then it has a future value 
$$
P_{t+1}=P_t\Big(1+\frac{r}{m}\Big)^{m\times n}.
$$
If compounding times  increases, then the future value will also rise. In case of continuous compounding we have
$$
P_{t+1}=P_t\lim_{m \rightarrow \infty}\Big(1+\frac{r}{m}\Big)^{m \times n}=P_t e^{r \times n}.
$$
If $n=1$ then continuous compounding is $P_{t+1}=P_te^{r},$ which we can present as 
$$
r=\log\Big(\frac{P_{t+1}}{P_t}\Big)
$$
the log return.  Hence that the log return of an asset is known as continuously compounded rate of return.
\subsection{Volatility as Measure of Risk}\label{intro_volatility}
\index{Volatility}
\index{Risk}
Volatility tells us, on average, how much value of an asset can go down or go up. The standard deviation calculated from the daily log returns are known as the \textbf{volatility} at daily levels, i.e.,

$$
\sigma =\sqrt{\mathbb{V}ar(r_{t})}.
$$
Suppose $r_t, r_{t-1},\cdots,r_{t-k+1}$ are $k$ single period log-return of an asset, where 
\begin{eqnarray*}
\mathbb{E}(r_{t-i+1})&=&\mu,~~\forall ~~i=1,2,\cdots,k,\\
\mathbb{V}ar(r_{t-i+1})&=&\sigma^2,~~\forall~~i=1,2,\cdots,k,\\
\mathbb{C}ov(r_{t-i+1})&=&0,~~\forall~~i\neq j=1,2,\cdots,k.
\end{eqnarray*}
That is the covariance matrix is 
\begin{eqnarray*}
\Sigma=\left(\begin{array}{cccc}
\sigma^2 & 0 & ... & 0 \\
0 & \sigma^2 & ... & 0 \\
\vdots  &  \vdots & ... & \vdots\\
0 & 0 & ... & \sigma^2  
\end{array}\right)_{k \times k}
\end{eqnarray*}
The $k$-period return can be presented in matrix notation as
\begin{eqnarray*}
r_{t}(k)&=&r_t+r_{t-1}+...+r_{t-k+1}\\
&=&c^T\mathbf{r},
\end{eqnarray*}
where $c^T=(1,1,...,1)_{k}$ is an unit vector of order $k$ and $\mathbf{r}=\{r_t,r_{t-1},...,r_{t-k+1}\}$. The mean and variance of the $k$-\emph{period return} are 
\begin{eqnarray*}
\mathbb{E}(r_{t}(k))&=& c^T\mathbf{\mu}=k\mu,\\
\mathbb{V}ar(r_{t}(k))&=&c^T\Sigma c =k\sigma^2
\end{eqnarray*}
Therefore $k$-\emph{period volatility} is $\sqrt{k}\sigma$.

\section{Efficient Market Hypothesis}\label{sec_EMH}

Economists and market analysts agree that if an arbitrage opportunity exists, everybody would like to follow that strategy which would thus disturb the equilibrium. \textbf{\emph{So going forward, a blanket assumption that arbitrage opportunities do not exist is being imposed}.} To check out more about \textbf{no-arbitrage opportunity}  in the market, see \cite{Shreve.2004.a,Shreve.2004.b}.

\subsection{Random Walk Hypothesis}
Let $\mathcal{F}_t=\sigma\{P_j:0\leq j \leq t\}$ is a $\sigma$-field generated by $\{P_0,\hdots,P_t\}$, where $P_t$ is the price at time point $t$. The condition of market equilibrium can be stated in terms of conditional expected returns on the basis of $\mathcal{F}_t$, where
$$
x_{t+1}=P_{t+1}-\mathbb{E}(P_{t+1}|\mathcal{F}_t)
$$
is the excess of market value at time $t+1$. It is the difference between the observed and the expected price that was projected at time $t$ on the basis of the information set $\mathcal{F}_t$. Then
$$
\mathbb{E}(x_{t+1}|\mathcal{F}_t)=0,
$$
which, by definition, says that the sequence $\{x_t\}$ is a ``fair game" with respect to the information sequence $\mathcal{F}_t$. Equivalently, let
$$
z_{t+1}=R_{t+1}-\mathbb{E}(R_{t+1}|\mathcal{F}_t),
$$
then, 
$$
\mathbb{E}(z_{t+1}|\mathcal{F}_t)=0
$$
so that the sequence $\{z_t\}$ is also ``fair game", with respect to information sequence $\{\mathcal{F}_t\}$. Let 
$$
\omega(\mathcal{F}_t)=\Big[\omega_1(\mathcal{F}_t),\omega_2(\mathcal{F}_t),\hdots,\omega_p(\mathcal{F}_t)\Big],
$$
where $\omega_j(\mathcal{F}_t)$ is the amount of funds available at time $t$ that are to be invested in the $j^{th}$ security, $j=1,2\hdots,p$. The total excess market value at $t+1$ is
$$
V_{t+1}=\sum_{j=1}^p\omega_j(\mathcal{F}_t)[R_{j,t+1}-\mathbb{E}(R_{j,t+1}|\mathcal{F}_t)],
$$
which has value
$$
\mathbb{E}(V_{t+1}|\mathcal{F}_t)=0,
$$
so that the sequence $\{z_t\}$ is also ``fair game", with respect to information sequence $\{\mathcal{F}_t\}$. Note one thing here,
$$
\mathbb{E}(R_{t+1}|\mathcal{F}_t)=\mathbb{E}\Bigg(\frac{P_{t+1}-P_{t}}{P_{t}}\bigg|\mathcal{F}_t\Bigg),
$$
can be expressed as
\begin{eqnarray}\label{RFR_eqn_expected_price1}
\mathbb{E}(P_{t+1}|\mathcal{F}_t)=[1+\mathbb{E}(R_{t+1}|\mathcal{F}_t)]P_{t}.
\end{eqnarray}
If we assume in (\ref{RFR_eqn_expected_price1}), that for all $t$ and $\mathcal{F}_t$,
\begin{eqnarray}\label{RFR_eqn_expected_price2}
\mathbb{E}(P_{t+1}|\mathcal{F}_t)\geq P_t,
\text{ or equivalently }
\mathbb{E}(R_{t+1}|\mathcal{F}_t)\geq 0,
\end{eqnarray}
that means the price sequence $\{P_t\}$ for security follows a submartingale with respect to the information sequence $\mathcal{F}_t$. If (\ref{RFR_eqn_expected_price2}) holds an equality (that is expected return and price changes are zero), then price sequence follows a martingale.
\index{martingale}

In the efficient market model, the statement that current price of a security ``fully reflects" available data is assumed to imply that successive price changes are independent. Also, it is usually considered continuous changes or returns are independent and identically distributed. Together the two hypothesis constitute the random walk model. 

We can make a common working assumption as the returns are mutually independent and identically distributed ($i.i.d$) random variables with mean $\mu$ and variance $\sigma^2$.  We see, for the log return, $$
1+R_t=\exp(r_t)\geq 0,
$$
which implies $R_t\geq -1$. This satisfies the condition of limited liability, i.e., possible maximum loss is the total investment.  In addition,
\begin{eqnarray*}
1+R_t(k)&=&(1+R_t)(1+R_{t-1})\hdots(1+R_{t-k+1}),\\
&=&\exp(r_t)\exp(r_{t-1})\hdots\exp(r_{t-k+1}),\\
&=&\exp(r_t+r_{t-1}+\hdots+r_{t-k+1}).
\end{eqnarray*}
So to sum of $k$ period log-returns yield $k$-period gross return. Now note that
$$
\frac{P_t}{P_{t-k}}=1+R_t(k)=\exp(r_t+r_{t-1}+\hdots+r_{t-k+1}).
$$
can be expressed as for $k=t$,
$$
P_t=P_0\exp(r_t+\hdots+r_1).
$$

Suppose $r_1,r_2,...r_t$  be i.i.d with mean $\mu$ and standard deviation $\sigma$ . Let $P_0$  be an arbitrary starting point and 
$$
P_t=P_0+r_1+r_2+\hdots+r_t,~~t\geq 1.
$$
The process $P_0,P_1,P_2,\hdots$  is \textbf{random walk} and $r_1,r_2,\hdots$  are corresponding steps of that random walk. The conditional expectation and variance of $P_t$  given  $P_0$ is $\mathbb{E}(P_t|P_0)=P_0+\mu t$  and $\mathbb{V}ar(P_t|P_0)=\sigma^2 t$ . The parameter $\mu$ is the \textbf{drift} and set an overall trend of the random walk. The parameter $\sigma$ is the volatility and controls how much it fluctuates around $P_0+\mu t$.  Since the standard deviation of $P_t$  given $P_0$  is $\sigma\sqrt{t}$ , as $t$ increases the range of variability in the process increases. This means at the $t = 0$ we know very little about where the random walk will be in the remote future compared to its current spot value.
\index{drift}
\index{volatility}

%
%
Therefore, if the log returns are assumed to be i.i.d. random variables, then the price of the stock or market index, denoted by the process $P=\{P_t:t \geq 0\}$,  is the exponential of random walk or also known as the geometric random walk.

\subsection{Test for Random Walk Hypothesis}

If the price of a stock follows the geometric random walk, then we can write the log-return as
$$
p_t=p_{t-1}+r_t,
$$
where $p_t=\log(P_t)$ and $r_t$ follows the same distribution with drift parameter $\mu$ and volatility parameter $\sigma^2$. The random walk is said to have unit root. To understand what this means, we should consider the AR(1) model (i.e., Auto-Regressive model with lag 1),
$$
p_t=\phi p_{t-1}+r_t
$$
where $\phi=1$. The generic AR(1) model can be presented as
\begin{eqnarray*}
p_t&=&\phi p_{t-1}+r_t\\
&=& \phi (\phi p_{t-2} + r_{t-1}) + r_t\\
&=& \phi^2 p_{t-2} + \phi r_{t-1} + r_t \\
&\vdots&\\
&=& \phi^k p_{t-k} + \phi^{k-1} r_{t-(k-1)}+\hdots + \phi r_{t-1} + r_t\\
&=& \phi^k p_{t-k} + \sum_{i=0}^{k-1}\phi^{i-1} r_{t-(i-1)}.
\end{eqnarray*}
\begin{itemize}
\item If $\phi = 1$ then the process is non-stationary. Because $\sum_{i=0}^{k-1}\phi^{i-1} r_{t-(i-1)}$ accumulates the information over time. \textbf{Hence a random walk is a non-stationary process}. 
\item However $|\phi|<1$, i.e., $-1<\phi<1$ implies, the process is stationary.
\item If $\phi=0$ that means the process is stationary and $p_t$ and $p_{t-1}$ are independent $\forall t$. 
\end{itemize}

We may ask here what a stationary process is? How does it look? How it looks different from the non-stationary process?

As the series $\{p_t: t \geq 0\}$ is a random walk (i.e., $\phi=1$) the incremental steps (i.e., log-returns) are independent and stationary process, we can write it as
$$
r_t=\phi_1 r_{t-1}+\epsilon_t
$$
where $\phi_1=0$ and $\epsilon_t$ is white noise with mean $\mu$ and variance $\sigma^2$.

In order to check if the price of a stock follows the geometric random walk, we have to check following three things.
\begin{enumerate}
\item First, we should check if $\{p_t\}$ is a non-stationary process, i.e.,
$$
p_t=\phi p_{t-1}+r_t;
$$
check if $\phi=1$ or $\phi<1$.
\item Second, we check if the log-returns are stationary process, i.e.,
$$
r_t=\phi_1 r_{t-1}+\epsilon_t;
$$
check if $\phi_1=1$ or $\phi_1<1$.
\item Second check only tells we if a log-returns are stationary, but it does not check if $\phi_1=0$ or not. In addition $\phi_1=0$ only implies pairwise independence. It does not check the mutual independence of $r_t$. So we should check if the serial correlations of $r_t$ are 0 or not. That is check if $\rho_1=\rho_2=\hdots=\rho_H=0$, where $\rho_h=corr(r_t,r_{t+h})$ is the lag $h$ auto-correlation.
\end{enumerate}
\subsection{Dickey-Fuller test for Stationarity in a Return Series}
A test involving much more narrowly-specified null and alternative hypotheses was proposed by \cite{dickey_fuller_test}. The test compares the null hypothesis
$$
H_0: p_t=p_{t-1}+r_t
$$
i.e., that the series is a random walk without drift, where $r_t$ is a white noise with mean 0 and variance $\sigma^2$. The alternative hypothesis is
$$
H_1: p_t=\mu + \phi p_{t-1} + r_t
$$
where $\mu$ and $\phi$ are constant with $|\phi|<1$. According to $H_1$, the process is stationary AR(1) with mean $\frac{\mu}{1-\phi}$. We implement the Dickey Fuller test using \texttt{adf.test} function in \texttt{tseries} package.

Note that first we have to check if the log prices are a random walk. Then we have to check if the log-returns are also a random-walk or if the log-return follows the stationary distribution. In these two cases, we can use Dickey-Fuller test. Then on the third step, we check if consecutive log-returns are independent or not!
\subsection{Ljung-Box test for independence in a Return Series}
We can check the independence of log-return between the consecutive days via Ljung-Box test (see, \cite{ljung_box_test}) for autocorrelation. Suppose the correlation between $r_t$ and $r_{t+h}$ is denoted as $\rho_h=corr(r_t,r_{t+h})$ and known as lag $h$  auto-correlation. The null hypothesis is $\rho_h = 0$ for. That is,
$$
H_0: \rho_1=\rho_2=\hdots=\rho_H=0 ~\forall~ t,
$$
\begin{center}vs.\end{center}
$$
H_1: \text{At least one inequality.}
$$
The test statistic for the Ljung-Box test is
$$
Q=n(n+2)\sum_{h=1}^H\frac{\hat{\rho}_h^2}{n-h},
$$
where $n$ is the sample size, $\hat{\rho}_h$ is the sample autocorrelation of lag $h$. We can show under $H_0$, $Q$ follows a chi-square distribution, $\chi^2_{(h)}$. The Ljung-Box test can be done in \texttt{R} using \texttt{Box.test} function available in \texttt{stats} package.

\subsection{Test for Efficient Hypothesis with \texttt{R}}

Market analysts always want to know if the market is efficient? Here we see; how we can check and test if the market is efficient. Since EMH is a ``hypothesis"; therefore we can run a statistical test to check whether EMH is true! We can consider the market index as proxy or representative of the market as a whole and check if the market index is following a random walk. If it follows a random walk, then it is good enough to claim that market is efficient.

\begin{figure}[ht]
    \centering
    \begin{tabular}{cc}
    \includegraphics[width=6cm]{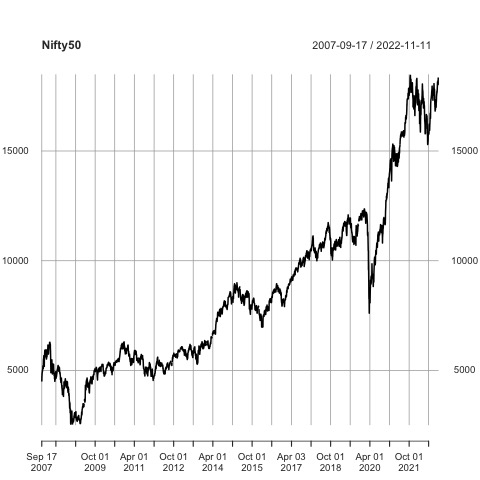}
         & \includegraphics[width=6cm]{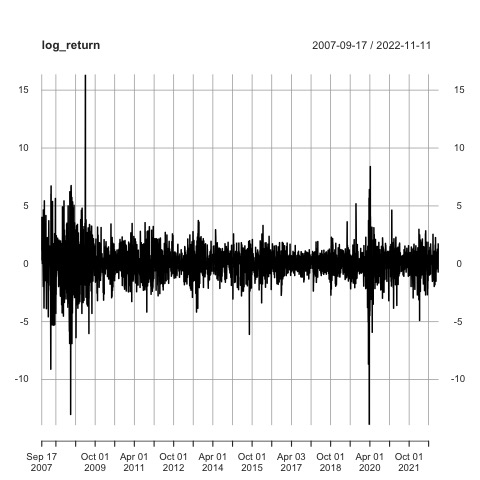}
     \\
     (a) Nifty 50 index values & (b) log return of Nifty 50\\
    \end{tabular}
    \caption{Values of Nifty 50 and log return over time. We consider the log-return as $\text{log-return}=\log(P_t/P_{t-1})\times 100$}
    \label{fig:Nifty_plot}
\end{figure}

We consider the adjusted close value of Nifty50 market index of National Stock Exchange. We can download data using the \texttt{R}-package the \texttt{quantmod}.  We consider the adjusted close value, and compute the log return of the Nifty. First, plot values of Nifty50 and log return over time in Figure (\ref{fig:Nifty_plot}).

\begin{tcolorbox}
Plot Nifty50 values and its log-return over time.
\begin{verbatim}
library(quantmod)
library(tseries)
## Download Nifty50 from Yahoo
getSymbols("^NSEI",src = "yahoo")
Nifty50 = NSEI$NSEI.Adjusted
plot(TCS)

### calculate and plot the log-return
log_return = diff(log(TCS))*100
plot(log_return)
\end{verbatim}
We presented both plot in the Figure (\ref{fig:Nifty_plot}).
\end{tcolorbox}

\noindent We try to answer the following questions to check if the Indian market is efficient.
\begin{enumerate}
    \item Are the values of Nifty50 non-stationary?
    \item Are the log-returns of Nifty50 non-stationary?
    \item Are the log-returns uncorrelated?
    \item Do the log-returns follow Gaussian distribution?
\end{enumerate}

\begin{tcolorbox}
Step 1: Check if values of Nifty 50 is non-stationary
\begin{verbatim}
## Augmented Dickey-Fuller (adf) test for unit-root
> library(tseries)
> adf.test(na.omit(Nifty50))
data:  na.omit(Nifty50)
Dickey-Fuller = -2.1463, Lag order = 15, p-value = 0.5164
alternative hypothesis: stationary
\end{verbatim}  
\textbf{Inference}: Fail to reject null hypothesis. That is Nifty 50 values are non-stationary.
\end{tcolorbox}

\begin{tcolorbox}
Step 2: Check if log-returns are non-stationary with Dickey-Fuller test
\begin{verbatim}
> adf.test(na.omit(log_return))

data:  na.omit(log_return)
Dickey-Fuller = -14.327, Lag order = 15, p-value = 0.01
alternative hypothesis: stationary
\end{verbatim}  
\textbf{Inference}: We reject the null hypothesis. That is log-returns of Nifty 50 are stationary.
\end{tcolorbox}

\begin{tcolorbox}
Step 3: Check if the log-returns are uncorrelated with Ljung-Box test.
\begin{verbatim}
> Box.test(log_return,lag=10,type = "Ljung-Box")
data:  log_return
X-squared = 37.234, df = 10, p-value = 5.155e-05
\end{verbatim}  
\textbf{Inference}: We reject null hypothesis as p-value is significantly small. That is log-returns of Nifty 50 are correlated.
\end{tcolorbox}

\begin{tcolorbox}
Step 4: Check if the log-returns are Normal with Shapiro-Wilk test for normality, (see \cite{shapiro_wil_test}).
\begin{verbatim}
## Shapiro-Wilk test for normality
## Null Hypothesis: log-return follows Normal distribution
## Alternative Hypothesis : log-return does not 
##                          follow a normal distribution
> shapiro.test(as.vector(log_return))
data:  as.vector(log_return)
W = 0.89425, p-value < 2.2e-16
\end{verbatim}  
\textbf{Inference}: We reject null hypothesis as p-value is significantly small. That is log-return of Nifty 50 does not follow Gaussian distribution.
\end{tcolorbox}

\begin{figure}[ht]
    \centering
    \begin{tabular}{cc}
    \includegraphics[width=6cm]{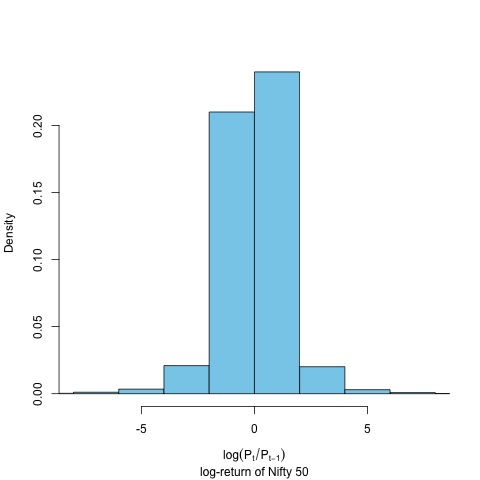}     & \includegraphics[width=6cm]{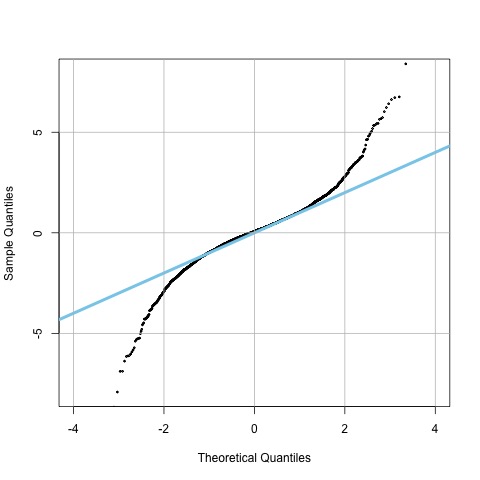} \\
    (a) Histogram of log-return & (b) qq-norm plot of log-return\\
    \end{tabular}
 
    \caption{(a) Histogram of log-return of Nifty50, and (b) The qq-norm plot of log-return indicates that the log-return of Nifty 50 certainly does not follow the Gaussian distribution.}
    \label{fig:TCS_ret_hist}
\end{figure}

\begin{tcolorbox}
We  draw histogram and qqplot of log-return and presented in Figure (\ref{fig:TCS_ret_hist}).
\begin{verbatim}
## Draw histogram of log-return of the FTSE
hist(log_return,main="",col="blue",nclass = 20,probability = TRUE)
qqnorm(log_return,xlim=c(-4,4)
       ,ylim=c(-4,4)
       ,cex=0.3)
abline(a=0,b=1,col="blue")
grid(col="red")
\end{verbatim}    
\end{tcolorbox}

In conclusion, the Indian stock market is not efficient, as the market returns are correlated and do not follow the Gaussian distribution. However, the log returns are stationary.

\subsection{Capital Asset Pricing Model}

As an investor, we would like to assess if the price of a stock is less than its expected level. If we see the stock is already overpriced, then the chance that it will appreciate further will be less, and we would like to sell the stock. Other investors would also like to sell the stock as we have the same information. The stock will fall back to its expected level. On the contrary, if the stock is underpriced, many investors would like to buy it with the assumption that the price will rise to its expected level. In finance, this is known as `Asset Pricing,' and corresponding mathematical models are known as  `Capital Asset Pricing Models' (CAPM). It explains the expected risk and returns for a single asset or portfolio. We can estimate the risk premium of assets using the CAPM.

Suppose $rp_{tj}=(r_{tj}-r_f)$ is excess return over the $r_f$ (i.e., risk free rate) for the $j^{th}$ stocks on  the $t^{th}$ day. The $rp_j$ is also known as \emph{risk premium} of the $j^{th}$ stock. We consider the model as follows:
\begin{equation}\label{eqn_CAPM}
\rb=\X\bfb + \bfep,
\end{equation}
where $\rb=((rp_{tj}))_{n\times P}$ is the matrix of risk-premium for $P$ many assets that are available in the market over $n$ days; $\X\bfb$ is the systematic return due to market index, where
$$
\X=((\boldmath{1},\rb_m))_{n\times 2},    
$$
is the design matrix with the first column being the unit vector or the place holder for intercept and the second
column being the risk-premium for the market index over the risk free rate $r_f$;
$$
\bfb = \begin{pmatrix}
    \alpha_1 & \alpha_2 & \cdots & \alpha_P\\
    \beta_1 & \beta_2 & \cdots & \beta_P \\
\end{pmatrix}_{2 \times P}.
$$
If the market is efficient then according to (see, \cite{sharpe_1964},\cite{Black1972}), then $\alpha_i = 0 ~~\forall i=1,2,\cdots,P$ and $\beta_i$ is the measure of systematic risk due to market movement; $\bfep=((\varepsilon_{tj}))_{n \times P}$ is the idiosyncratic return of the asset. In general one can consider a $k$-factor model with $\X$ being $n\times (k+1)$ dimensional, where the first column of $\X$ is constant and other columns are all suitable factors. The coefficients $\bfb$ is $(k+1)\times P$ dimensional and we shall denote it as
$$
\bfb = (\theta_1,\cdots,\theta_P),~~~\text{where}~~~\theta_i=(\alpha_i,\beta_i,b_i^{(1)},\cdots,b_i^{(k-1)}),~~i=1,2,\cdots,P.
$$
\begin{remark}
    Equation (\ref{eqn_CAPM}), for the $k$ factor model, if a portfolio is constructed based on $\tilde{P}$ many assets all of which have $\alpha=0$, $\beta=1$ and $b^{(j)}=0 \forall j \in \{1,2,\cdots,k-1\}$, then the portfolio return will mimic the market return.
\end{remark}
The covariance of $\rb_t$ is $\bfS$ which can be decomposed into
\begin{equation}
    \label{eqn_cov_matrix}
    \bfS = \bfb^T\bfS_{X}\bfb + \bfS_{\bfep},
\end{equation}
where $\bfS_{\bfep}=diag(\sigma_1^2,\sigma_2^2,\cdots,\sigma_P^2)$ and $\bfS_{X}$ is the covariance matrix of $\X$. In the next section we develop portfolio optimisation using the CAPM strategy.

\section{Portfolio Risk Analysis}\label{sec_portfolio_risk_analysis}

\subsection{Portfolio Selection by Minimising Idiosyncratic Risk}
Let us consider a portfolio $\bfo=\{\omega_1,\omega_2,\cdots,\omega_P\}$ where $\omega_i \geq 0$, $i=1,\cdots,P$, $\sum_{i=1}^{P}\omega_i=1$. Markowitz's portfolio optimization (see, \cite{markowitz_1952}) can be expressed as the following quadratic programming problem:
\begin{equation}
\label{eqn_markowitz_optim}
\min_{\bfo} ~\bfo^T\bfS\bfo, \text{ subject to } \bfo^T\mathbf{1}_P=1 \text{ and } \bfo^T\bfm = \mu_k.   
\end{equation}
Here , $\mathbf{1}_P$ is a $P$-dimensional vector with one in every entry and $\mu_k$ is the desired level of return. The portfolio covariance can be decomposed into two parts as,
\begin{eqnarray*}
   \bfo^T \bfS \bfo &=& \bfo^T [ \bfb^T\bfS_X\bfb + \bfS_{\bfep}]\bfo\\
   &=& \bfo^T \bfb^T\bfS_X\bfb \bfo + \bfo^T \bfS_{\bfep} \bfo, 
\end{eqnarray*}
where first part explains the portfolio volatility due to market volatility and the second part explains portfolio volatility due to idiosyncratic behaviour of the stock. We assume $\sigma_i^2$'s are bounded $\forall i$. 
\begin{result}\label{reslt_diversification}
    Under the CAPM model (\ref{eqn_CAPM}), covariance matrix (\ref{eqn_cov_matrix}) and assumption (\ref{eqn_markowitz_optim}), if $P\longrightarrow \infty$, and $M_{\omega:P}=\max\{\bfo\}\longrightarrow 0 $ and $\sigma_{\max}^2=\max \{\bfS_{\bfep}\}<\infty$, then
    $$
    \lim_{P\longrightarrow 0} \bfo^T \bfS_{\bfe} \bfo =0
    $$
\end{result}
\begin{proof}
We consider, 
\begin{eqnarray*}
\bfo^T \bfS_{\bfe} \bfo &=& \sum_{i=1}^{P} \omega_i^2 \sigma_i^2 \\    
&\leq & \sigma_{\max}^2 \sum_{i=1}^{P}\omega_i^2,~~\sigma_{\max}^2=\max \{\bfS_{\bfep}\},\\
&\leq& \sigma_{\max}^2 M_{\omega:P}\sum_{i=1}^{P}\omega_i, ~~~M_{\omega:P}=\max\{\bfo\},\\
&=& \sigma_{\max}^2 M_{\omega:P}.
\end{eqnarray*}
Clearly, if $P\longrightarrow \infty$ and $M_{\omega:P}\longrightarrow 0$ $\implies \bfo^T \bfS_{\bfe} \bfo \longrightarrow 0$. 
\end{proof}
\begin{remark}
    The Result (\ref{reslt_diversification}) is based on real analysis and not probabilitic result. A detailed discussion about the Result (\ref{reslt_diversification}) can be found in \cite{das_sen_2020}.
\end{remark}

\begin{result}\label{remark_equal_weight}
     If $P$ is fixed, then $M_{\omega:P}\geq \frac{1}{P}$. If we want to minimise $M_{\omega:P}$, then the we must have \textbf{equal weights portfolio}, i.e., $M_{\omega:P}=\frac{1}{P}$. So 
     $$
     \text{if } M_{\omega:P}=\frac{1}{P}, \text{ then } \bfo^T \bfS_{\bfe} \bfo
     $$ is minimum.
\end{result}

\begin{remark}
    The idiosyncratic risk of the portfolio will be washed out, as the portfolio's size increases and the portfolio's maximum weight is bounded. The portfolio's performance will be a function of the systematic risk explained by the market indices.
\end{remark}
\begin{remark}
   The above result mathematically demonstrates the proverb, "\textbf{one should not put all eggs in one basket.}"
\end{remark}
\begin{remark}
    Thus we can select $\tilde{P}(\ll P)$ many assets for the portfolio (out of $P$ many assets available in the market) such that the idiosyncratic risk becomes negligible, i.e., $\forall \delta >0, \exists \tilde{P}_{\delta}$ such that for $\tilde{P}>\tilde{P}_{\delta}$,
    \begin{eqnarray*}
        \bfo^T_{\tilde{P}} \bfS_{\tilde{P}}\bfo_{\tilde{P}}<\delta;
    \end{eqnarray*}
    and portfolio return is mostly explained by $\theta$ only. Note that here $\tilde{P}$ is the effective size of the portfolio.
\end{remark}
\begin{remark}
    \textbf{Oracle Set}: Suppose the market is not efficient and there are $q$ many assets whose $\alpha>0$, where $q<\tilde{P}\ll P.$ Let us call this set as $A_q$. We can construct a portfolio with $\tilde{P}$ many assets, such that 
    \begin{equation}\label{eqn:oracle_set}
        \mathbb{P}(A_q \subset B_{\tilde{P}})\geq 1 - \eta,
    \end{equation}
    where $B_{\tilde{P}}$ is the set of assets in the portfolio, $0\leq \eta \leq 1$ and $\bfo^T_{\tilde{P}} \bfS_{\tilde{P}}\bfo_{\tilde{P}}<\delta$. 
\end{remark}
\begin{example}
    Suppose the market consists of $P=2000$ stocks and a portfolio manager wants to build the portfolio with $\tilde{P}=100$ stocks. If $q=5$ many stocks are available  with $\alpha_j>0,~j=1,2,3,4,5$, then the portfolio manager would like to build a portfolio, such that $A_{q=5}$ is the subset of manager's selected portfolio $B_{\tilde{P}=100}$. In other words, the manager wants to build her portfolio in such a way that she does not want to miss out the set of five under-valued stocks $A_{q=5}$. That is, she wants to employ a statistical methodology, where $\mathbb{P}(A_{q=5}\subset B_{\tilde{P}=100})$ would be very high. Note that if the market is efficient, then $A_q$ will be a null set.
\end{example}
The problem reduces to identifying the oracle set $A_q$. Essentially, it is a multiple testing problem, where we select those stocks in the portfolio $B_{\tilde{P}}$ for which we reject the following null hypothesis:
$$
H_{0i}: \theta_i = \theta_0 ~~~vs.~~~H_{1i}:\theta_i \neq \theta_0,~~~i=1,2,,\cdots,P.
$$
where $\theta_0=(0,1,0,\cdots,0)$. \cite{das_sen_2020} presented optimal test such that Equation (\ref{eqn:oracle_set}) is satisfied.

\subsection{Regularising Portfolio Risk Analysis}\label{sub_sec_regularise_risk}

It is important to estimate the volatility and determine the primary sources of volatility. Often the number of assets of well-diversified mutual funds or pension funds is more than thousands. However, portfolio managers are concerned about the stationarity in long time-series data. They are interested only in recent volatility, which considers daily returns of a month and sometimes even less. As the number of assets ($P$)in a portfolio is greater than the number of days of return ($n$), the rank of the covariance matrix is less than complete; such cases yield non-unique solutions. Generally, it is known as the ``ill-posed" problem. 

So for large portfolio, as $P\longrightarrow \infty$ and $n\longrightarrow \infty$, however $P/n = \lambda,$ where $0<\lambda<1$ is a constant; the sample covariance matrix
$$
\mathbf{S} =\frac{1}{n-1}\sum_{i=1}^{n}(\rb_i-\bar{\rb})(\rb_i-\bar{\rb})^T,
$$
where $\rb_i=(r_{i1},\cdots,r_{iP})^T$, $\bar{\rb}=(\bar{r}_1,\cdots,\bar{r}_P)^T$, and $\bar{r}_j=\frac{1}{n}\sum_{i=1}^{n}r_{ij}$, $j=1,2,\cdots,P$. Note that $rank(\mathbf{S})=\min(n,p)=n<p$, where $\mathbf{S}$ is a $p\times p$ matrix. 
In such cases, there are two problems. First, as  the sample covariance matrix $\mathbf{S}$ is less than full rank, the sampling distribution of $\mathbf{S}$ is degenerate. Hence no valid statistical inference can be implemented with such $\mathbf{S}$. Second, as we try to implement Markowitz's portfolio optimization, as described in Equation (\ref{eqn_markowitz_optim}), the portfolio covariance matrix $\bfS$ is unknown. However, we cannot estimate it with $\mathbf{S}$, as  $\mathbf{S}$ is not a full rank. The Markowitz optimization (\ref{eqn_markowitz_optim}) would not have any unique solution, if we use $\mathbf{S}$.
To address this issue one can use the shrinkage estimator for covariance (see, \cite{Ledoit_Wolf_2003}) for portfolio optimization. However, the sample distribution of these estimators are not completely understood. Hence one cannot run the statistical inference using these shrinkage estimators. \cite{Das_Halder_Dey_2013} and \cite{Das_Dey_2010} presented the Bayesian inference for covariance matrix, particularly when $\mathbf{S}$ is less than full rank matrix.
\cite{Das_Halder_Dey_2013} considered that inverse Wishart prior for $\bfS$, while $\mathbf{S}$ follow Wishart distribution, i.e.,
\begin{eqnarray*}
\bfS &\sim& \mathcal{W}^{-1}(n_0,\bfP),\\
\mathbf{S}&\sim& \mathcal{W}(n-1,\bfS),    
\end{eqnarray*}
where $\bfP$ is a postive definite matrix and  the posterior distribution of $\bfS$ is
$$
\bfS|\mathbf{S}\sim \mathcal{W}^{-1}(n_0+n-1,\bfP+\mathbf{S}).
$$
If $n<P$, then we choose the prior degrees of freedom as
$$
n_0=(P-n)+c,
$$
where $c>0$. This ensures posterior distribution to be proper. The posterior mode of $\bfS$ is
$$
M(\bfS|\mathbf{S})=q\frac{\bfP}{n_0+P+1}+(1-q)\frac{\mathbf{S}}{n-1},
$$
where $q=\frac{n_0+p+1}{n_0+n+p}$. The posterior mode of $\bfS$ is a shrinkage estimator, a weighted average of prior distribution's mode and sample covariance estimator. The advantage of \cite{Das_Halder_Dey_2013} is that full posterior distribution of $\bfS$ is known. Hence we can run full Bayesian inference on $\bfS$. In addition, since posterior mode of $\bfS$ is positive definite and full rank, we can run portfolio optimization and further portfolio risk analysis.

\subsection{Bayesian Inference with Regularised Covariance}

The portfolio return over the period is
$$
r_p =\rb\bfo,
$$
where $r_p=\{r_{p1},\cdots,r_{pn}\}^T$, $\rb=((rp_{tj}))_{n\times P}$ is the risk-premium matrix, and $\bfo=\{\omega_1,\cdots,\omega_P\}^T$, such that $\omega_j$ is the $j^{th}$ asset's weight. The portfolio volatility is defined as
$$
\sigma_P =\sqrt{\bfo^T\bfS\bfo}.
$$
The portfolio weights $\bfo$, and the covariance structure of the portfolio plays a crucial role as regulators of the portfolio's total volatility $\sigma_P$. However, it is also essential  to quantify how sensitive the portfolio volatility is concerning a slight change in $\bfo$. We can achieve it by differentiating the volatility with respect to weight, and it is known as the 'Marginal Contribution to Total Risk' (MCTR), see \cite{Menchero_2011} and \cite{Baigent_2014}.
\begin{Defn}
    The `Marginal Contribution to Total Risk' (MCTR) is defined as
    $$
    \frac{\partial(\sigma_P)}{\partial \bfo}=\frac{1}{\sigma_P}.\bfS.\bfo = \bfvr,
    $$
    where $\bfvr=\{\varrho_1,\cdots,\varrho_P\}$, . The MCTR for asset $i$ is 
    $$
    \varrho_i=\frac{1}{\sigma_P}\sum_{j=1}^P \sigma_{ij}\omega_j.
    $$    
\end{Defn}
\begin{Defn}
    The `Conditional Contribution to Total Risk' (CCTR) is the amount that an asset that contributes to the total portfolio volatility. In other words, if $\zeta_j=\omega_j\varrho_j$ is the CCTR of the asset $j$ then 
    $$
    \sigma_P=\sum_{j=1}^P\zeta_j = \sum_{j=1}^P \omega_j\varrho_j.
    $$
    Therefore the total volatility is weighted average of the MCTR.
\end{Defn}

\begin{remark}
 A regularized estimate of $\bfS$ is required to estimate the MCTR and CCTR, while the weights are fixed. However, for all practical purposes we are interested in estimation of $\mathbb{P}(\zeta <0)$ or  $\mathbb{P}(\varrho <0)$. Because negative CCTR or MCTR of an asset means that asset actually reduces the volatility.   
\end{remark}
In Section (\ref{sub_sec_regularise_risk}), we presented  that the posterior distribution of $\bfS$ follows $\mathcal{W}^{-1}(n_0+n-1,\bfP+\mathbf{S}).$ To estimate the contribution to risk, we present the following Monte Carlo algorithm.

\RestyleAlgo{ruled}
\begin{algorithm}
\caption{Bayesian Monte Carlo Algorithm to Estimate Contribution to Risk}\label{algo_bayes_mc}
\For{$i = 1 : N$}{
     $\bfS^{(i)} \gets \mathcal{W}^{-1}(n_0+n-1,\bfP+\mathbf{S})$\;
     $\sigma_P^{(i)} \gets \sqrt{\bfo^T\bfS^{(i)}\bfo}$\;
     $\bfvr^{(i)} \gets \frac{1}{\sigma_P^{(i)}}\bfS^{(i)}\bfo$\;
     $\zeta_j^{(i)} \gets \varrho_j^{(i)}\omega_j,~~\forall j=1,\cdots,P$\;
}
\end{algorithm}
\begin{remark}
In the Algorithm (\ref{algo_bayes_mc}), each iterations are independent. Hence parallel implementation of the algorithm is very simple. In fact we can consider the algorithm to be an `embarrassingly parallel', see \cite{Matloff_2011}.    
\end{remark}

\begin{remark}
    Implementing the algorithm in parallel might not be required if $P$ is small. However, as $P$ is large, generating $\bfS^{(i)}$ will be slow for a large portfolio. Thus, consequent computation in all the other steps will also be slow. In such cases, parallelization of the algorithm improves the time performance of the algorithm. 
\end{remark}
\noindent The MC estimate of CCTR and other risk metrics can be estimated, like 
$$
\mathbb{P}(\zeta_j > 0 )=\frac{1}{N}\sum_{i=1}^{N}\mathbb{I}(\zeta_j^{(i)}>0),
$$
once the MC samples are generated.

\subsection{Analysing the Extreme Risk of Portfolio}

The volatility risk is a measure of  average risk of the portfolio. However, it does not say anything about the risk of large losses, or extreme risk. \textbf{Value at Risk} (VaR) of a portfolio  is a measure of extreme risk. It states that the portfolio will lose more than a large amount is the $\alpha$ quantile of the portfolio return, i.e., 
$$
VaR_{\alpha}(r_P)=-\inf\{v: \mathbb{P}(r_p<v)\leq \alpha\}=F_{r_P}^{-1}(\alpha),
$$
where $v$ is the loss of the portfolio, and $\alpha \in (0,1)$ is the confidence level.

\begin{example}
If a portfolio of stocks has a one-day $1\%$ VaR of \rupee 1 crore, there is $1\%$ probability that the portfolio will decline in value by more than \rupee 1 crore over the next day. 
\end{example}

\begin{remark}
The VaR provides a measure of how much extreme financial risk we are exposed to. It provides a structured methodology for critically thinking about risk, and consolidating risk across an organization. VaR can be applied to individual stocks, portfolios of stocks, gold, and other commodity, etc.
\end{remark}

\begin{remark}
    One has to have a distributional assumption about $F$. Instead of assuming a particular parametric distribution, we considered empirical distribution. 
\end{remark}
\noindent The VaR is a frequency measure. It does not measure the expectation of the amount lost. The \textbf{Expected Shortfall} (ES), aka., the Conditional VaR or CVaR, is the measure of risk of expected amount of large loss, i.e.,
$$
ES_{\alpha}=\mathbb{E}\big(r_p|r_p <-VaR_{\alpha}(r_p)\big)=\int_{-\infty}^{-VaR_{\alpha}}r_pdF(r_p).
$$
\begin{remark}
 A \textbf{coherent risk measure}  satisfies four properties, i.e., (i) monotonicity, (ii) sub-additivity, (iii) homogeneity, and  (iv) translational invariance. The VaR is not a coherent risk measure. However, ES is a coherent risk measure. It makes ES a more desirable risk measure than VaR. 
\end{remark}

\section{Nonparametric Bootstrap Methods in Risk Analysis}

Nonparametric Bootstrap statistics is an algorithmic approach that typically employs a simple random sample with replacement (SRSWR) scheme. It belongs to the broader category of resampling strategies. Bootstrap was introduced by \cite{}. Despite its apparent simplicity, the concept revolutionised statistics by replacing analytical derivations with brute computational power. Moreover, in cases where parametric assumptions are known to be incorrect, Nonparametric Bootstrap provides an appropriate solution. For example, in the Capital Asset Pricing Model (CAPM), the underlying distribution is often assumed to be Gaussian, which is not accurate, as evidenced by Figure (\ref{fig:TCS_ret_hist}).

The nonparametric bootstrap method involves resampling returns from a given set of asset returns. Suppose \( \rb = \{r_1, r_2, \cdots, r_n \} \) represents the log-returns in the original sample from a distribution \( F(\cdot) \), and \( T_n = T_n(r_1, r_2, \cdots, r_n) \) is a statistic that estimates a parameter \(\theta\). The sampling distribution of \( T_n \) depends on \( F(\cdot) \). \emph{The core idea of the bootstrap is to estimate the cumulative distribution function (cdf) \( F(\cdot) \) using the empirical cdf \( F_n(\cdot) \)}. The empirical cdf \( F_n(\cdot) \) is the nonparametric maximum likelihood estimate (MLE) of the cdf \( F(\cdot) \). Bootstrapping based on \( F_n(\cdot) \) is known as the nonparametric bootstrap. We can draw samples from \( F_n(\cdot) \), which is equivalent to drawing independent and identically distributed (iid) samples from \( \{r_1, r_2, \cdots, r_n\} \). This process can be repeated as many times as needed.

\subsection{Bootstrap Framework}

Since \( F(\cdot) \) is unknown, the sampling distribution of \( T_n \) is also unknown. As a result, we cannot determine the variance of \( T_n \), i.e., \( \text{Var}(T_n) \), nor the confidence interval of \( T_n \), i.e., \( \text{CI}(T_n) \). Resample $\rb_{nb}^*=\{r_1^*,r_2^*,\cdots,r_n^*\}_b$ from $\rb_n$ using SRSWR scheme; $b=1,2,\cdots,B$. For each resample $b$, we can compute $T_{nb}^*$; $b=1,2,\cdots,B$. Then we can compute:
\begin{eqnarray*}
\bar{T}_n^B &=& \frac{1}{B}\sum_{b=1}^{B}T_{nb}^*; ~~~~Var(T_n)^B=\frac{1}{B}\sum_{b=1}^{B}(T_{nb}^*-\bar{T}_n^B)^2\\
CI(T_n)^B&=&\{T_n+G_B^{-1}(\alpha/2)\sqrt{Var(T_n)^B},\\
&& ~~~~~~~~T_n+G_B^{-1}(1-\alpha/2)\sqrt{Var(T_n)^B}\},
\end{eqnarray*}
where $\frac{T_{nb}^*-T_n}{\sqrt{Var(T_n)^B}}\sim G_B$. Due to SLLN, one can show, as $B \longrightarrow \infty$
\begin{eqnarray*}
\bar{T}_n^B &\longrightarrow&  T_n~~\text{almost surely};\\
Var(T_n)^B &\longrightarrow& Var(T_n)~~\text{almost surely}\\
CI(T_n)^B  &\longrightarrow& CI(T_n)~~\text{almost surely},
\end{eqnarray*}
\begin{eqnarray*}
G^B &\longrightarrow& F_{T_n}(\cdot)~~\text{in law.}\\
\end{eqnarray*}

\subsection{Residual Bootstrap Regression for CAPM}

Consider the capital asset pricing model
$$
\rb_n=\X_{n\times p}\bfb_p+\bfe_n,
$$
where $\mathbb{E}(\bfe)=0$, $\mathbb{V}ar(\bfe)=\sigma^2\I_n$, and $\bfe \stackrel{iid}{\sim} F(\cdot)$, $F(\cdot)$ is unknown cdf. The OLS estimatoris $\hat{\bfb}_n=(\X^T\X)^{-1}\X^T\rb$; and $\mathbb{V}ar(\hat{\bfb}_n)=\sigma^2(\X^T\X)^{-1}$. The residuals are $\bfe=\rb-\X\hat{\bfb}_n$  or $\epsilon_i=y_i-\x_i^T\hat{\bfb}_n,~~i=1,2,\cdots,n$. Suppose $F_n(\cdot)$ is the empirical cdf of $\bfe$. $\bfe^*_b \stackrel{iid}{\sim}F_n$ (i.e., $\bfe^*_b$ is resampled from $\bfe$ using SRSWR), $b=1,2,\cdots,B$. We calculate,
$$
\rb^*_b=\X\hat{\bfb}_n+\bfe^*_b,
$$
then estimate resample coefficients $\hat{\bfb}_{n:b}^*$ as
\begin{eqnarray*}
\hat{\bfb}_{n:b}^*&=&(\X^T\X)^{-1}\X^T\rb^*_b\\
&=& \hat{\bfb}_n + (\X^T\X)^{-1}\X^T\bfe^*_b\\
\text{where }
\mathbb{E}(\hat{\bfb}_{n:b}^*)&=&\hat{\bfb}_n.
\end{eqnarray*}
Then we have the bootstrap estimate,
  $\bar{\bfb}_B=\frac{1}{B}\sum_{b=1}^{B}\hat{\bfb}_b^*$.
and the bootstrap variance is 
$$
\mathbb{V}ar(\bar{\bfb}_B)=\frac{1}{B}\sum_{b=1}^{B}(\hat{\bfb}_b^*-\bar{\bfb}_B)^2.
$$

\subsection{Paired Bootstrap Regression}

We consider the model
$$
  \rb_n=\X_{n\times p}\bfb_p+\bfe_n,
$$
where $\mathbb{E}(\bfe)=0$, $\mathbb{V}ar(\bfe)=\bfS$, and $(r_i,\x_i) \stackrel{iid}{\sim} F(\cdot)$, where $F(\cdot)$ is an unknown cdf.
Suppose $\{(r_i^*,\x_i^*),i=1,2,...n\}_b=\mathcal{D}_b$ are iid samples from empirical $F_n(\cdot)$, where $b=1,2,\cdots,B$. The estimates of $\bfb$ from $b^{th}$ resample is,
$$
\hat{\bfb}_b^*=(\X^{*T}_b\X^{*}_b)^{-1}\X^{*T}_b\rb_b^{*}.
$$
The bootstrap estimate is $\bar{\bfb}_B=\frac{1}{B}\sum_{b=1}^{B}\hat{\bfb}_b^*$, and the bootstrap variance is 
$$
\mathbb{V}ar(\bar{\bfb}_B)=\frac{1}{B}\sum_{b=1}^{B}(\hat{\bfb}_b^*-\bar{\bfb}_B)^2.
$$
\begin{remark}
If the residuals are heteroscedastic, the paired bootstrap remains a consistent estimator. However, when the residuals are heteroscedastic, the residual bootstrap is not a consistent estimator.  
\end{remark}

\subsection{CAPM with Bootsrap Statistics usin \texttt{R}}
Now, we will demonstrate the Capital Asset Pricing Model using the nonparametric bootstrap regression technique in \texttt{R}. First, we will download the data for Reliance and Nifty 50 from Yahoo. Next, we will calculate the log-returns and then the risk-premium. After that, we will fit the CAPM using the OLS method and obtain the residuals.
\begin{tcolorbox}
\begin{verbatim}
library(tseries)
start_date<-"2024-01-01"
end_date<-"2024-06-30"

rel<-get.hist.quote(instrument = "RELIANCE.NS",start=start_date
            ,end=end_date,quote="AdjClose"
            ,provider = "yahoo",quiet = TRUE)

nifty<-get.hist.quote(instrument = "^NSEI",start=start_date
        ,end=end_date,quote="AdjClose"
        ,provider = "yahoo",quiet = TRUE)

data <-merge(nifty,rel)
rt<-diff(log(data))
risk_free_rate<-0.06/252

\end{verbatim}
\end{tcolorbox}

\begin{tcolorbox}
\begin{verbatim}
## risk premium
rt<-rt-risk_free_rate

## Fit CAPM using OLS
CAPM <- lm(Adjusted.rel~Adjusted.nifty,data=rt)

## Extract residual and fitted values
resid <- CAPM$residuals
y_hat <- CAPM$fitted.values

\end{verbatim}
\end{tcolorbox}

\begin{tcolorbox}
\begin{verbatim}
set.seed(6587)
rt1<-data.frame(rt)
n <- nrow(rt1)
B<-1000
beta_star<-matrix(NA,nrow=B,ncol = 2)
colnames(beta_star)<-c('alpha','beta')
R.squred_star.pair<-rep(NA,B)

for(b in 1:B){
  id_star<-sample(1:n,n,replace = TRUE)
  rt_star<-rt1[id_star,]
  
  CAPM_star<-lm(Adjusted.rel~Adjusted.nifty,data=rt_star)
  sum_star<-summary(CAPM_star)
  beta_star[b,] <- coef(CAPM_star)
  R.squred_star.pair[b] <- sum_star$adj.r.squared
}

\end{verbatim}
\end{tcolorbox}

\begin{tcolorbox}
\begin{verbatim}
sum_boot <-cbind(apply(beta_star,2,mean)
                 ,apply(beta_star,2,sd)
                 ,apply(beta_star,2,quantile,probs=0.025)
                 ,apply(beta_star,2,quantile,probs=0.975))
colnames(sum_boot)<-c('Estimate','Std.Error','2.5%','97.5%')
cat('OLS Estimates of alpha and beta')
ols_estimates<-coefficients(summary(CAPM))
rownames(ols_estimates)<-c('alpha','beta')
round(ols_estimates,4)
cat('Paired Bootstrap Estimates of alpha and beta')
round(sum_boot,4)

Paired Bootstrap Estimates of alpha and beta

      Estimate Std.Error    2.5%  97.5%
alpha   0.0005    0.0010 -0.0012 0.0024
beta    1.2439    0.1309  0.9804 1.5023

    \end{verbatim}
\end{tcolorbox}

\begin{figure}
    \centering
    \includegraphics[width=12cm,height=8cm]{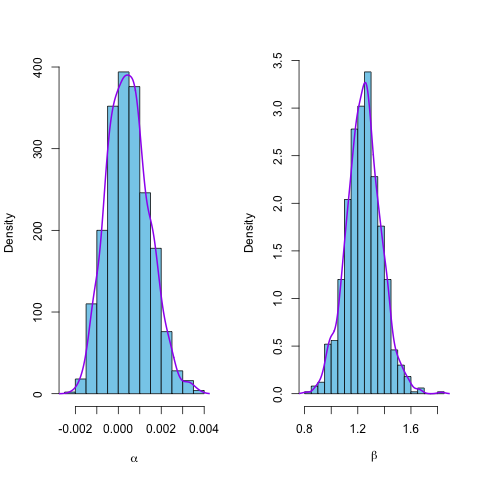}
    \caption{Bootstrap histogram of $\alpha$ and $\beta$ of CAPM.}
    \label{fig:enter-label}
\end{figure}

\begin{tcolorbox}
\begin{verbatim}
par(mfrow=c(1,2))
hist(beta_star[,'alpha'],main = '',col='skyblue'
        ,freq = FALSE,xlab=expression(alpha),nclass = 20)
lines(density(beta_star[,'alpha']),col='purple',lwd=2)
hist(beta_star[,'beta'],main = '',col='skyblue',freq = FALSE
        ,xlab=expression(beta),nclass = 20)
lines(density(beta_star[,'beta']),col='purple',lwd=2)
\end{verbatim}
\end{tcolorbox}

\begin{tcolorbox}
\textbf{Residual Bootstrap Regression}

\begin{verbatim}
set.seed(6587)

ols_resid <- CAPM$residuals
ols_pred  <- CAPM$fitted.values
Adjusted.nifty <- rt$Adjusted.nifty
B<-1000
beta_star2<-matrix(NA,nrow=B,ncol = 2)
colnames(beta_star2)<-c('alpha','beta')
R.squred_star.resid<-rep(NA,B)
n <- nrow(rt1)
for(b in 1:B){
  id_star<-sort(sample(1:n,n,replace = TRUE))
  resid_star<-ols_resid[id_star]
  pred_star <- ols_pred+resid_star
  CAPM_star<-lm(pred_star~Adjusted.nifty)
  sum_star<-summary(CAPM_star)
  beta_star2[b,] <- coef(CAPM_star)
  R.squred_star.resid[b] <- sum_star$adj.r.squared
}

\end{verbatim}
\end{tcolorbox}

\begin{tcolorbox}
\textbf{Summary of Residual Bootstrap Regression for CAPM}
\begin{verbatim}
sum_boot2 <-cbind(apply(beta_star2,2,mean)
                  ,apply(beta_star2,2,sd)
                  ,apply(beta_star2,2,quantile,probs=0.025)
                  ,apply(beta_star2,2,quantile,probs=0.975))
colnames(sum_boot2)<-c('Estimate','Std.Error','2.5%','97.5%')
ols_estimates<-coefficients(summary(CAPM))
\end{verbatim}
\end{tcolorbox}

\begin{tcolorbox}
\textbf{OLS Estimates of $\alpha$ and $\beta$}

\begin{verbatim}
> rownames(ols_estimates)<-c('alpha','beta')
> round(ols_estimates,4)
      Estimate Std. Error t value Pr(>|t|)
alpha   0.0006     0.0010  0.6285   0.5309
beta    1.2441     0.0981 12.6794   0.0000
\end{verbatim}
\end{tcolorbox}

\begin{tcolorbox}
\textbf{Residual Bootstrap Estimates of $\alpha$ and $\beta$}
\begin{verbatim}
> round(sum_boot2,4)
      Estimate Std.Error    2.5%  97.5%
alpha   0.0006    0.0010 -0.0012 0.0026
beta    1.2462    0.0838  1.0887 1.4216

\end{verbatim}
\end{tcolorbox}

\begin{tcolorbox}
\textbf{
Paired Bootstrap Estimates of $\alpha$ and $\beta$}
\begin{verbatim}
> round(sum_boot,4)
      Estimate Std.Error    2.5%  97.5%
alpha   0.0006    0.0010 -0.0012 0.0026
beta    1.2379    0.1283  0.9935 1.5086  

\end{verbatim}
\end{tcolorbox}
\section{Empirical Risk Analysis of Passive Investment}\label{sec_empirical_evidence}

\subsection{Philosophy of Passive Investment}
The philosophy of passive investment strategy yields from the `efficient market hypothesis.' The logic is that since the market is efficient, it does not make sense that anyone will be able to beat the market consistently for a long time. Therefore, investing in funds that mimic the market only makes sense. It resulted in the popularisation of the exchange-traded fund  (ETF), where the fund invests by mimicking the index weights. However, those weights are based on market capitalisation and are not necessarily optimal for investors. What kind of passive investment strategy would be better? In Result (\ref{remark_equal_weight}), we demonstrated that for a large portfolio with a fixed $P$ number of assets, an equal weight portfolio would yield minimum idiosyncratic risk, while we have no control over the systematic risk due to market movement.

\subsection{Comparing of Two Passive Investment Strategies }

In Section (2), we demonstrated that the Indian stock market based on the Nifty50 index is not efficient. So it raises the question how it affects the risk profile of the passive strategy (i.e., portfolio weight). To check this question, we decided to consider three different portfolios. First, we consider the passive investment portfolio with Nifty weights as the portfolio weights. Second, we consider Markowiz's portfolio weights optimised with regularised covariance matrix. Third, we consider the equal weight as discussed in the Result (\ref{remark_equal_weight}), an alternate passive investment strategy.

\begin{figure}
    \centering
    \begin{tabular}{cc}
    \includegraphics[width=7cm]{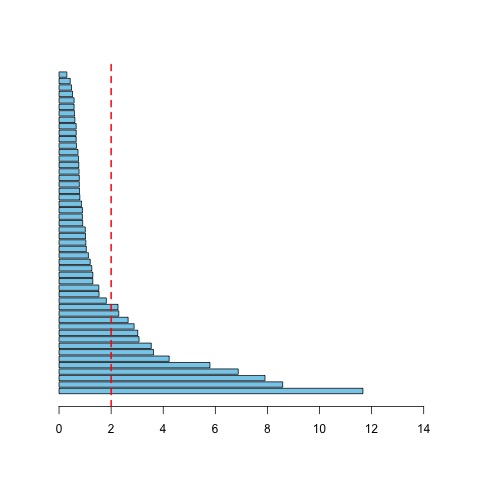}
         & \includegraphics[width=7cm]{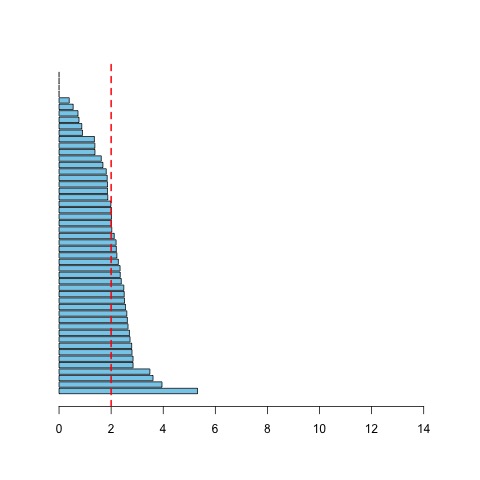}
     \\
     (a) Nifty Weights & (b) Markowitz's Optimised Weights
    \end{tabular}
    \caption{The weights of Nifty 50 are much more skewed than Markowitz's optimised weights. The horizontal red-dash line represents the equal-weight portfolio. Markowitz's optimised weights are comparatively less deviate from equal weights.}
    \label{fig:my_label}
\end{figure}

\begin{table}[ht]
\centering
\begin{tabular}{lrrr}
  \hline
 & Portfolio with  & Portfolio with  & Portfolio with  \\ 
 &Nifty Weights & Markowitz Weights& Equal Weights\\
  \hline
Before the War & 11.35 & 8.30 & 9.03 \\ 
  During the War & 13.66 & 11.41 & 11.30 \\ 
   \hline
\end{tabular}
\caption{Portfolio Volatility before the Russia-Ukraine war and during the war. Portfolio with Nifty weights have uniformly higher volatility than other portfolios, before and during the war. However, the volatility risk profile is portfolio with equal weights are that of similar to portfolio with Markowitz's optimal weights.}\label{tabl_portfolio_volatility}
\end{table}

Out of the three portfolios we considered here, the passive investment strategy with Nifty weights remained the same before and during the war. Similarly, the equal portfolio weights remained the same before and during the war. We use the before-war returns for Markowiz's portfolio weights to estimate the weights using Markowitz's optimisation with regularised covariance. Then we use the same weights to calculate the portfolio volatility during the war.

In Table (\ref{tabl_portfolio_volatility}), we present the portfolio Volatility before the Russia-Ukraine war and during the war. The portfolio with Nifty weights has uniformly higher volatility than other portfolios before and during the war. However, the volatility risk profile of a portfolio with equal weights is similar to a portfolio with Markowitz's optimal weights because Markowitz's weights are close to equal weights. The Result (3.2) states that idiosyncratic risk would be minimum for an equal weights portfolio. Therefore if we really want to have a passive investment strategy, we should try to implement equal-weight portfolios than Nifty weights. As Nifty weights are based on market capitalisation and not close to equal weights, a portfolio with Nifty 50 weights would contain a significant idiosyncratic risk than an equal weights portfolio.

\begin{table}[ht]
\centering
\begin{tabular}{lrrr}
  \hline
 & Portfolio with &Portfolio with & Portfolio with \\
 & Nifty Weights & Equal Weights & Markowitz Weights \\ 
 & VaR (ES) & VaR (ES) & VaR (ES) \\
  \hline
Before the War & -5.75 (-6.07) & -5.53 (-5.84)& -3.74 (-3.85)\\ 
  During the War & -6.70 (-7.03) & -6.11 (-6.23) & -6.34 (-6.62)\\ 
   \hline
\end{tabular}
\caption{Portfolio VaR and Expected Shortfall (ES) before the Russia-Ukraine war and during the war. The numbers inside the parenthesis are ES and outside are VaR. The Portfolio with Nifty weights have uniformly higher VaR and ES than other portfolios, before and during the war.}
\end{table}

\section{Conclusion}\label{sec_conclusion}

In this work, we present that if we want to be passive investors, we should follow an equal-weight portfolio strategy instead of investing in Exchange Traded Fund like NETF, which mimics the Nifty50. We also present an analysis of how portfolios perform to idiosyncratic events like the Russian invasion of Ukraine. We found that the equal weight portfolio has a uniformly lower risk than the Nifty 50 portfolio before and during the Russia-Ukraine war. We also showed in Results (\ref{reslt_diversification},\ref{remark_equal_weight}) that if we push the maximum weights of the portfolio towards the equal weight portfolio, then the idiosyncratic risk of the portfolio is minimal. As a result, the equal-weight portfolio has a uniformly lower risk before and during the Russia-Ukraine war than the Nifty 50 portfolio.

\bibliographystyle{plainnat}
\bibliography{Biblio_DataBase}

\begin{thebibliography}{16}
\providecommand{\natexlab}[1]{#1}
\providecommand{\url}[1]{\texttt{#1}}
\expandafter\ifx\csname urlstyle\endcsname\relax
  \providecommand{\doi}[1]{doi: #1}\else
  \providecommand{\doi}{doi: \begingroup \urlstyle{rm}\Url}\fi

\bibitem[Baigent(2014)]{Baigent_2014}
G.~G. Baigent.
\newblock X-sigma-rho and market efficiency.
\newblock \emph{Journal of Economic and Financial Studies}, 2, 2014.

\bibitem[Black(1972)]{Black1972}
Fischer Black.
\newblock Capital market equilibrium with restricted borrowing.
\newblock \emph{The Journal of Business}, 45:\penalty0 444--455, 1972.

\bibitem[Das and Dey(2010)]{Das_Dey_2010}
Sourish. Das and Dipak.~K. Dey.
\newblock On bayesian inference for generalized multivariate gamma
  distribution.
\newblock \emph{Statistics and Probability Letters}, 51:\penalty0 1492--1499,
  2010.

\bibitem[Das and Sen(2020)]{das_sen_2020}
Sourish Das and Rituparna Sen.
\newblock Sparse portfolio selection via bayesian multiple testing.
\newblock \emph{Sankhya - B}, Nov 2020.

\bibitem[Das et~al.(2017)Das, Halder, and Dey]{Das_Halder_Dey_2013}
Sourish. Das, Aritra. Halder, and Dipak.~K. Dey.
\newblock Regularizing portfolio risk analysis: A bayesian approach.
\newblock \emph{Methodology and Computing in Applied Probability}, 19:\penalty0
  865--889, 2017.

\bibitem[Dickey and Fuller(1979)]{dickey_fuller_test}
David.~A. Dickey and Wayne.~A. Fuller.
\newblock Distribution of the estimators for autoregressive time series with a
  unit root.
\newblock \emph{Journal of the American Statistical Association}, 74:\penalty0
  427--431, 1979.

\bibitem[Exchange(2022)]{NETF}
National~Stock Exchange.
\newblock Netf: Tata nifty exchange traded fund.
\newblock \url{https://www.nseindia.com/get-quotes/equity?symbol=NETF}, 2022.
\newblock Accessed: 2022-11-18.

\bibitem[Ledoit and Wolf(2003)]{Ledoit_Wolf_2003}
O.~Ledoit and M.~Wolf.
\newblock Improved estimation of the covariance matrix of stock returns with an
  appli- cation to portfolio selection.
\newblock \emph{Journal of Empirical Finance}, pages 603--621, 2003.

\bibitem[Ljung and Box(1978)]{ljung_box_test}
G.~M. Ljung and George.~P. Box.
\newblock On a measure of lack of fit in time series models.
\newblock \emph{Biometrika}, 65:\penalty0 297--303, 1978.

\bibitem[Markowitz(1952)]{markowitz_1952}
Harry. Markowitz.
\newblock Portfolio selection.
\newblock \emph{Journal of Finance}, pages 77--91, 1952.

\bibitem[Matloff(2011)]{Matloff_2011}
N.~Matloff, editor.
\newblock \emph{The Art of R Programming: A Tour of Statistical Software
  Design.}
\newblock The Journal of Portfolio Management, 2011.
\newblock ISBN 9781593274108: 347.

\bibitem[Menchero and Davis(2011)]{Menchero_2011}
Jose. Menchero and Ben. Davis.
\newblock Risk contribution is exposure times volatility times correlation:
  Decomposing risk using the x-sigma-rho formula.
\newblock \emph{The Journal of Portfolio Management}, 37:\penalty0 97--106,
  2011.

\bibitem[Shapiro and WILK(1965)]{shapiro_wil_test}
S.~S. Shapiro and M.~B. WILK.
\newblock An analysis of variance test for normality (complete samples).
\newblock \emph{Biometrika}, 52:\penalty0 591--611, 1965.

\bibitem[Sharpe(1964)]{sharpe_1964}
William.~F. Sharpe.
\newblock Capital asset prices: A theory of market equilibrium under conditions
  of risk.
\newblock \emph{Journal of Finance}, 19:\penalty0 425--442, 1964.

\bibitem[Shreve(2004{\natexlab{a}})]{Shreve.2004.a}
Steven~E. Shreve, editor.
\newblock \emph{Stochastic Calculus for Finance I The Binomial Asset Procing
  Model}.
\newblock Springer, 2004{\natexlab{a}}.
\newblock ISBN 13:987-0387-24968-1.

\bibitem[Shreve(2004{\natexlab{b}})]{Shreve.2004.b}
Steven~E. Shreve, editor.
\newblock \emph{Stochastic Calculus for Finance II Continuous Time Model}.
\newblock Springer, 2004{\natexlab{b}}.
\newblock ISBN 978-0-387-40101-0.

\end{thebibliography}

\section*{Appendix A: No Arbitrage: No Free Lunch}\label{chap-No-Free-Lunch}

\subsection*{Portfolio of Bonds and Shares}

A \emph{bond} \index{Bond} is a debt security, earning a fixed rate of interest $r$ in each unit of time. If we make an investment $B_0$ at time $0$ in the bond is worth $B_0(1+r)^k$ at time $k$. If our interest earning is compounding over $k$-period, then the bond will be worth $B_0(1+r)^{k}$ at time $k$. As bond can be bought or sold, as an investor we can invest or borrow at the rate of interest $r$. We can trade \emph{shares} of the \emph{stock}  of a specific company in the stock market. The price $P_k$ at which one share of a stock can be traded is modeled as a stochastic process. \emph{Bond} and \emph{stock} are together known as securities (aka. \emph{primary securities}.) We can consider our investment in \emph{house} or \emph{real estate} as primary security. However, the market for the real estate behaves very differently than the typical bond and stock market. The quantitative finance primarily focuses on the bond and stock market and anything related to that.

\noindent Suppose price of a stock is such that
$$
\mathbb{P}(P_1\geq P_0(1+r))=1,~~\mathbb{P}(P_1 > P_0(1+r))>0.
$$
Then we can borrow an amount $P_0$ and buy one share of the stock at time 0. At time 1, we can sell the stock at a price $P_1$, settle our debt by paying $P_0e^r$ and our profit $Profit=P_1-P_0(1+r)$ is non-negative with probability one and is strictly positive with positive probability. It can be argued if such stock is available in the market and all information is available to everyone, then many investors like us would like to invest large amounts of money (by borrowing) into the stock; since there is nothing to lose and something to be gained. This will disturb the equilibrium and push the price of the stock (at time 0) up. Such opportunities are known as an \emph{arbitrage opportunity}. In general, in a market consisting of several securities, an \emph{arbitrage opportunity} is a strategy of buying and selling the securities without any investment, such that it leads to profit (strictly positive) with positive probability without any risk of a loss. 

The price of the stock on the $k^{th}$ day is denoted by $P_k$. The price $P_0$ of the stock on day zero is assumed to be deterministic, $P_0=p_0$. Also the face value of the bond is 1 on the morning of day zero. For $k \geq 0$, let $\theta_k^{S}$ denote the number of shares we decide to hold on the morning of $k^{th}$ day, before the market opens. Suppose $\theta_k^{B}$ denote the number of bonds we choose to keep. If for $k \geq 1$, $\theta_k^{S} \geq \theta_{k-1}^{S}$, we buy $\theta_k^{S}-\theta_{k-1}^{S}$ shares and if $\theta_k^{S} < \theta_{k-1}^{S}$, then we sell $\theta_{k-1}^{S}-\theta_{k}^{S}$ shares. We have same interpretation for bonds. In order to implement a strategy we may have to put in extra money on certain days while  surplus on other days. However, we  consider a special kind of strategy, known as \emph{self-financing strategies}. These are trading strategies where there is no money put in and there is no surplus  on any day except for the initial investment $x$, where
\begin{equation}\label{eqn_chap1_initial_investment}
x=\theta_0^B+\theta_0^SP_0.
\end{equation}
Thus, on a given day, we only moves our money from shares to bonds or vice-versa. The shares and bonds held by us is known as our \textbf{\emph{portfolio}}.

\subsection*{Martingle and Arbitrage}
Let $\mathcal{F}_i=\sigma\{P_j:0\leq j \leq i\}$ is a finite $\sigma$-field generated by $\{P_0,\hdots,P_i\}$. Suppose there exists a probability measure $\mathbb{Q}$ on $\mathcal{P}$ such that $\{P_i(1+r)^{i},\mathcal{F}_i\}$ is a $\mathbb{Q}$-martingale and
$$
\mathbb{Q}(p_0,\hdots,p_N)>0~~~\forall~(p_0,\hdots,p_N)\in \mathcal{P}.
$$
For any self-financing strategy $\eta$, the worth of the portfolio on $k^{th}$ day is
$$
V_k(\eta)(P_0,\cdots,P_N)
$$
is a $\mathbb{Q}$-martingale, and hence
\begin{equation}\label{eqn_chap1_Q_martingale}
\mathbb{E}^{\mathbb{Q}}[V_k(P_0,\hdots,P_N)(1+r)^{-k}]=V_0(\eta).
\end{equation}
Let $\eta$ be such that $V_0(\eta)=0$, that is initial investment is zero, then (\ref{eqn_chap1_Q_martingale}) implies
$$
V_N(\eta)(p_0,\hdots,p_N)=0~~\forall (p_0,\hdots,p_N)\in \mathcal{P}.
$$
Thus arbitrage opportunities, i.e., 
\begin{equation}\label{eqn_chap1_arbitrage_def1_cond1}
V_N(\eta)(p_0,\hdots,p_N) \geq 0, ~\mbox{ for all } (p_0,\hdots,p_N) \in \mathcal{P},
\end{equation}
and
\begin{equation}\label{eqn_chap1_arbitrage_def1_cond2}
V_N(\eta)(p_0^*,\hdots,p_N^*) > 0, ~\mbox{ for some } (p_0^*,\hdots,p_N^*) \in \mathcal{P}.
\end{equation}
strategies satisfying (\ref{eqn_chap1_arbitrage_def1_cond1}) and (\ref{eqn_chap1_arbitrage_def1_cond2}), do not exist.
\begin{tcolorbox}
\begin{theorem}\label{Thm_cap1_Fundamental_asset_pricing_thm}\textbf{Fundamental Theorem of Asset Pricing}\\
The following statements are equivalent.
\begin{enumerate}
\item \textit{No arbitrage}.
\item There eqxists a probability measure $\mathbb{Q}$ on $\mathcal{P}$ such that $\{P_i(1+r)^{i},\mathcal{F}_i\}$ is a $\mathbb{Q}$-martingale and
$$
\mathbb{Q}(p_0,\hdots,p_N)>0~~~\forall~(p_0,\hdots,p_N)\in \mathcal{P}.
$$
\end{enumerate}
\end{theorem}    
\end{tcolorbox}

\subsection*{Asset Pricing Model\label{Chap_intro_section_BAPM}}
So far we assumed that there is an underlying stochastic process for stock price movement. But no effort is rendered to model the stock price movement. We need a useful probability model to use as a computationally tractable approximation to theoretical models. The model should explain no-arbitrage pricing and its relation to risk-neutral pricing. Also, the model should incorporate the theory of conditional expectation and martingale theory which is the core of the risk-neutral pricing. To fulfill all these purposes the \textbf{Asset Pricing Model,} (APM) serves a good point to start. With this motive in mind, we present the geometric Brownian motion (GBM) model which is a slightly different from that usually found in practice.

\begin{figure}[ht]
    \centering
    \includegraphics{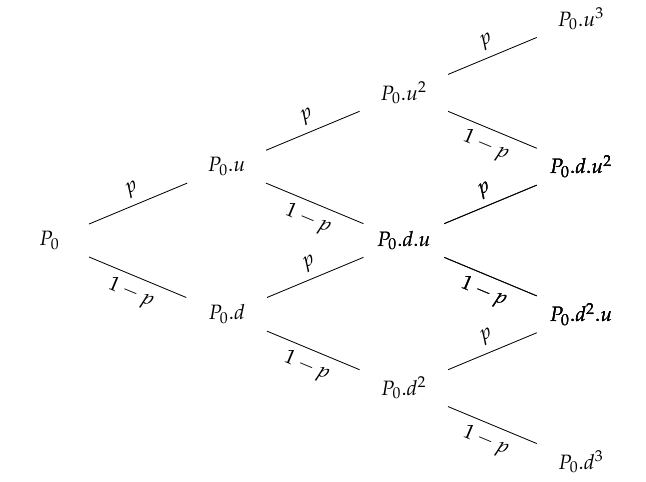} 
    \caption{Three Period Binomial Model}
    \label{Fig_Chap1_BAPM}
\end{figure}

We can model the price movement over time interval $[0,t]$ as a sequence of multiple binary steps. We can break the range into $n$ equal intervals per unit time. So over the range, the stock price will move through $nt$ binary steps and let the price follow one of the up or down branches each over each subinterval as presented in the figure (\ref{Fig_Chap1_BAPM}). Note that we should choose $n$ and $t$ in such a way such that $nt$ to be an integer. 

\noindent \textbf{Assumption}:
\begin{enumerate}
\item Each step is independent of previous one.
\item Parameters $p$, $u$ and $d$ are same for each step.
\end{enumerate}
Suppose $P_0$ is the initial stock price. At each time step, the stock price either goes up by a factor of $u$ with probability $p$ or down by a factor of $d$ with probability $1-p$. We define the following indicator variable as 
\begin{equation*}
X_i=\bigg\{\begin{array}{cc}
1 & \text{ if } P_i=P_{i-1}u,\\
0 & \text{ if } P_i=P_{i-1}d.
\end{array}
\end{equation*}
The fundamental asset pricing theorem (\ref{Thm_cap1_Fundamental_asset_pricing_thm}) states that to have the no arbitrage opportunity, there must be a probability of these outcomes that satisfy the $\mathbb{Q}$-martingale measure conditions at each step. That is
$$
\mathbb{E}(P_i|P_{i-1})=P_{i-1}(1+\frac{r}{n}),
$$
where
$$
pP_{i-1}u+(1-p)P_{i-1}d=P_{i-1}(1+\frac{r}{n}),
$$
which implies
$$
\hat{p}=\frac{(1+\frac{r}{n})-d}{u-d}.
$$
Note that $\hat{p}$ is known as the risk-neutral probability. The stock price at time $t$ is still a function of initial price $P_0$ and result of $nt$ binary steps. Suppose
$$
U_{nt}:\text{ The number of up moves},
$$
$$
D_{nt}:\text{ The number of down moves},
$$
where
$$
nt=U_{nt}+D_{nt}.
$$
The random walk $R_{nt}$ is the difference between the number of up and down moves, i.e.,
$$
R_{nt}=U_{nt}-D_{nt}.
$$
The three steps model is graphically presented  in Figure (\ref{Fig_Chap1_BAPM}).

\subsection*{Geometric Brownian Motion}
Let $\sigma>0$ is a known constant and $r \geq 0$ is the interest rate. In this model the interest rate per period is $\frac{r}{n}$, the up factor is $u_n=\exp\{\sigma/\sqrt{n}\}$ and the down factor is $d_n=\exp\{-\sigma/\sqrt{n}\}$. The risk-renutral probability is then
$$
\hat{p}_n=\frac{1+\frac{r}{n}-\exp\{-\sigma/\sqrt{n}\}}{\exp\{\sigma/\sqrt{n}\}-\exp\{-\sigma/\sqrt{n}\}}.
$$
Let $t$ be an arbitrary positive rational rational number, and for each positive integer $n$, for which $nt$ is an integer, define
$$
R_{nt:n}=\sum_{i=1}^{nt}X_{k:n},
$$
where $X_{1:n},X_{2:n},\hdots,X_{n:n}$ are independent, identically distributed random variables with
$$
\tilde{\mathbb{P}}(X_{n:k}=1)=\hat{p}_n,~~~\tilde{\mathbb{P}}(X_{n:k}=-1)=1-\hat{p}_n,~~k=1,2,\hdots,n.
$$
The stock price at time $t$ in this model is
\begin{eqnarray}\label{eqn_chap1_Bapm_GBM}
P_n(t)&=&P_0u_n^{\frac{1}{2}(nt+R_{nt:n})}d_n^{\frac{1}{2}(nt-R_{nt:n})}\\
&=&P_0\exp\bigg\{\frac{\sigma}{2\sqrt{n}}(nt+R_{nt:n})\bigg\}\exp\bigg\{-\frac{\sigma}{2\sqrt{n}}(nt-R_{nt:n})\bigg\}\nonumber\\
&=&P_0\exp\bigg\{\frac{\sigma}{\sqrt{n}}R_{nt:n}\bigg\}.\nonumber
\end{eqnarray}
In the Chapter 3 of \cite{Shreve.2004.b}, it is presented that as 
\begin{eqnarray*}
n\rightarrow \infty, ~~\frac{\sigma}{\sqrt{n}}R_{nt:n}\stackrel{\mathcal{L}}{\longrightarrow}N\big((r-\sigma^2/2)t,\sigma^2t\big).
\end{eqnarray*}
Hence the following theorem can be presented as follows.
\begin{theorem}
As $n\rightarrow \infty$, the distribution of $P_n(t)$ in equation \ref{eqn_chap1_Bapm_GBM} converges to the distribution of 
\begin{equation}\label{eqn_chap1_GBM_risk_neutral}
P_t=P_0\exp\bigg\{\bigg(r-\frac{1}{2}\sigma^2\bigg)t+ \sigma W_t\bigg\},
\end{equation}
where $W_t$ is standard normal variable with 0 mean and variance $t$.
\end{theorem} 
The stochastic process $P=\{P_t : t \geq 0\}$ is a Geometric Brownian Motion (GBM) with drift parameter $\mu$ and volatility parameters $\sigma$, where
\begin{equation}\label{eqn_chap1_GBM_defn}
P_t=P_0\exp\bigg\{\bigg(\mu-\frac{\sigma^2}{2}\bigg)t + \sigma W_t \bigg\},
\end{equation}
where $W_t\sim N(0,t^2)$ is standard Brownian motion (BM). 
\begin{tcolorbox}
\textbf{Geometric Brownian Motion under Risk-Neutral Measure}\\

If we compare the equation (\ref{eqn_chap1_GBM_risk_neutral}) and (\ref{eqn_chap1_GBM_defn}), it is clear that under risk-neutral probability measure, if stock price follow the GBM then the drift parameter is the risk-free interest rate $r$. That is 
\begin{equation*}
P_t=P_0\exp\bigg\{\bigg(r-\frac{\sigma^2}{2}\bigg)t + \sigma W_t \bigg\}.
\end{equation*}

\end{tcolorbox}

\section*{Exercise}
\begin{enumerate}
    \item How does the concept of the time value of money affect investment decisions, and what are the key differences between the present value (PV) and future value (FV) of a cash flow?
    \item You are expected to receive \rupee 10,00,000 five years from now. If the annual discount rate is 6\%, what is the present value of this future amount?
    \item Given the following monthly closing prices of a stock for the first quarter of the year, calculate the quarterly return using the log-return method:
    \begin{enumerate}
        \item January: \rupee 120
        \item February: \rupee 130
        \item March: \rupee 125
    \end{enumerate}
    \item Explain the Efficient Market Hypothesis (EMH) and discuss its three forms. How does each form of EMH impact an investor's ability to achieve above-average returns?
    \item How does the Random Walk Hypothesis relate to the Efficient Market Hypothesis, and what are the implications of the Random Walk Hypothesis for technical analysis and investment strategies?
    \item What statistical tests can be used to evaluate the Random Walk Hypothesis in financial markets, and how do these tests determine if a time series of stock prices follows a random walk?
    \item Explain the Capital Asset Pricing Model (CAPM) and its key assumptions. How is the expected return of a security calculated using CAPM, and what is the significance of the $\alpha$ and $\beta$ coefficient in this model?
    \item Explain the concepts of portfolio risk and portfolio volatility. How are these measures calculated? Additionally, define Value at Risk (VaR) and describe its significance in portfolio management. How can VaR be calculated for a given portfolio?
    \item Explain the concepts of Marginal Contribution to Total Risk (MCTR) and Conditional Contribution to Total Risk (CCTR) in portfolio management. How are these metrics used to assess the risk contributions of individual assets within a portfolio? Provide the formulas for calculating MCTR and CCTR.
    \item Why is bootstrap statistics important in conducting risk analysis? Explain the advantages of using the bootstrap method over traditional parametric methods in estimating the risk measures for a portfolio.
    \item What is the difference between residual and paired bootstrap regression?

\end{enumerate}

\end{document}